\documentclass[useAMS,usenatbib]{mn2e}

\usepackage{amsmath,color}
\usepackage{graphicx}
\newcommand{\mnras}{MNRAS}
\newcommand{\apj}{ApJ}

\newcommand{\aj}{AJ}
\newcommand{\apjl}{ApJL}
\newcommand{\aap}{A\&A}

\newcommand{\pasp}{PASP}
\newcommand{\bain}{Bull.\,Astron.\,Inst.\,Neth.}

\title[Galactoseismology and the Local Density of Dark Matter]
{Galactoseismology and the Local Density of Dark Matter}
\author[Nilanjan Banik, Lawrence M. Widrow, and Scott
Dodelson]
{Nilanjan Banik$^{1,2}$, Lawrence M. Widrow$^{3}$, and Scott Dodelson$^{2,4}$\\
$^{1}$ Department of Physics, University of Florida, Gainesville, FL 32611, USA\\
$^{2}$ Fermi National Accelerator Laboratory, Batavia, IL 60510-0500\\
$^{3}$ Department of Physics, Engineering Physics \& Astronomy, Queen's
University, Kingston, ON K7L 3N6, Canada\\
$^{4}$ Kavli Institute for Cosmological Physics, Enrico Fermi
Institute, and the
Department of Astronomy \& Astrophysics, \\University of Chicago, Chicago, IL 60637}

\begin{document}

\date{in original form August, 2016}

\pagerange{\pageref{firstpage}--\pageref{lastpage}} \pubyear{2002}

\maketitle

\label{firstpage}

\begin{abstract}

  We model vertical breathing mode perturbations in the Milky Way's
  stellar disc and study their effects on estimates of the local dark
  matter density, surface density, and vertical force.  Evidence for
  these perturbations, which involve compression and expansion of the
  Galactic disc perpendicular to its midplane, come from the SEGUE,
  RAVE, and LAMOST surveys.  We show that their existence may lead to
  systematic errors of $10\%$ or greater in the vertical force
  $K_z(z)$ at $|z|=1.1\,{\rm kpc}$.  These errors translate to $\ga
  25\%$ errors in estimates of the local dark matter density.  Using
  different mono-abundant subpopulations as tracers offers a way out:
  if the inferences from all tracers in the Gaia era agree, then the
  dark matter determination will be robust. Disagreement in the
  inferences from different tracers will signal the breakdown of the
  unperturbed model and perhaps provide the means for determining the
  nature of the perturbation.

\end{abstract}

\maketitle

\section{Introduction}

In his seminal work on the vertical structure of the Galaxy,
\citet{oort1932} introduced a method to determine the gravitational
force perpendicular to the Galactic plane (the vertical force) near
the Sun from stellar kinematics.  Though Oort's main interest was in
developing a dynamical model for the Galaxy, he recognized that a
measurement of the vertical force as a function of distance from the
midplane could be combined with estimates of the density in visible
matter to infer the existence of unseen ``dark matter''.  To be sure,
his concept of dark matter was not what it is today.  Nevertheless,
the {\it Oort problem}, as efforts to determine the vertical structure
of the Milky Way have come to be known, provides our best
astrophysical handle on the {\it local} density of dark
matter\footnote{Estimates of the local dark matter density are
  sometimes referred to as the {\it Oort limit} though Oort limit may
  also refer to the outer edge of the Oort cloud.  On the other hand,
  {\it Oort problem} can also refer to the discrepancy between the age
  of star clusters in the solar neighbourhood and theoretical
  predictions for their disruption time.}.

The Oort problem relies on astrometric observations of stars that act
as tracers of the gravitational potential.  A key assumption is that
the tracers are in dynamical equilibrium with respect to their
vertical motions \citep{bahcall1984a, bahcall1984b, bienayme1987,
  kuijken1989a, kuijken1989b, kuijken1989c, kuijken1991, holmberg2000,
  holmberg2004, bovy2012, garbari2012, bovy2013}; for a recent review,
see \citet{read2014}. The assumption that the Galaxy is a steady state system dates back to Jeans (1922) in his critique of Kapteyn's Galactic model (Kapteyn 1922) and was central to Oort's analysis of the Galaxy's vertical structure.  This assumption is plausible since a typical
star will have completed many oscillations through the Galactic
midplane over its lifetime.  The discoveries of bulk vertical motions
in the stellar disc \citep{widrow2012, williams2013, carlin2013} and a
North-South asymmetry in the number counts of solar neighbourhood
stars \citep{widrow2012, yanny2013} call into question this
assumption.  In particular, the bulk motion observations imply that
the disc is undergoing compression and expansion perpendicular to the
midplane, in essence, a localized breathing mode.  Depending on its
phase, the breathing mode may manifest itself as a correlation between
the mean vertical velocity of the tracers and distance from the
midplane.  Indeed, the observations mentioned above suggest that
variations in the mean velocity with $z$ are of order $4-8\,{\rm
  km\,s}^{-1} {\rm kpc}^{-1}$.  Perturbations of this type can be
caused by the passage of a globular cluster, dwarf galaxy, or dark
matter sub-halo through the disc plane \citep{widrow2012, gomez2013,
  widrow2014, feldman2015} or by gravitational effects of a passing
spiral arm \citep{faure2014, debattista2014}.

In this paper, we investigate the impact of a breathing mode
perturbation on efforts to determine the local vertical force and dark
matter density.  If the entire stellar disc participates in a
breathing mode, then the surface density of stars within a particular
distance from the Galactic midplane, and therefore the vertical force,
will change with time.  Furthermore, if a tracer population
participates in a breathing mode, then models that treat it as an
equilibrium system will yield erroneous results for the inferred
vertical force.

It is common practice to use $K_{1.1}$, the magnitude of the vertical
force $1.1\,{\rm kpc}$ above and below midplane of the disk at the
Sun's position, as a dynamical constraint on the structure of the
Galaxy.  Much closer to the midplane and baryons will dominate the
vertical gravitational force.  Much further from the midplane and halo
stars will contaminate the sample of tracers.  As we will see, a
breathing mode perturbation that is consistent with the observed bulk
motions changes $K_{1.1}$ by only $\sim 1\%$.  \citep{widrow2012,
  read2014}.  On the other hand, the errors induced in estimates of
$K_{1.1}$ by using a similarly perturbed tracer population can be
$\sim 10\%$ or greater.

The usual strategy in the Oort problem is to find a solution to the
time-independent collisionless Boltzmann equation (CBE) that is
consistent with kinematic data for the tracers.  The analysis is
particularly simple when one assumes not only that the tracers are in
equilibrium, but that variations across the disc plane in the
gravitation potential and tracer distribution function (DF) can be ignored
and that the tracers are isothermal with respect to their vertical
velocities.  The first of these assumptions implies that the
gravitational potential $\psi(z)$ depends only on the total surface
density within a distance $z$ of the midplane.  That is
\begin{equation}
K_z(z) ~\equiv~ \left |\frac{\partial\psi}{\partial z}\right | ~=~ 2\pi G\Sigma(z)
\end{equation}
where $K_z(z)$ is the magnitude of the vertical acceleration and
$\Sigma$ is the total surface density between $-z$ and $z$.  The
second assumption implies that the vertical velocity dispersion
$\sigma$ of the tracers is independent of $z$.  With these two
assumptions, the CBE, or alternatively, the Jeans equation
perpendicular to the disc, together with the Poisson equation, imply
that
\begin{equation}
K_z ~=~ -\sigma^2\, \frac{\partial \ln{n}}{\partial z}
\end{equation}
where $n=n(z)$ is the number density of tracers.

The effects of variations across the disc plane in both the
gravitational potential and tracer DF are often viewed as corrections
to the plane-symmetric CBE, Jeans, and Poisson equations.
\citet{bovy2013} proposed a more rigorous method to model the full
gravitational potential.  The starting point in their analysis is to
sort stars into subpopulations selected for their helium and iron
abundance ratios.  These mono-abundant subpopulations (MAPs) are
treated as independent tracers of the gravitational potential and
modeled by the three-integral, quasi-isothermal DF of
\citet{binney2010, binney2011} and \citet{ting2013}.  The analysis
provides an estimate for the surface density and gravitational
potential as a function of $z$ and Galactocentric cylindrical radius
$R$.  

The expectation in \citet{bovy2013} is that all MAPs lead to the same
inferred gravitational potential within the model uncertainties.  In
this paper, we explore the converse, namely that variations in the
inferred force with $\sigma$ may reveal the presence of a breathing
mode perturbation.  We take the basic idea of using MAPs from
\citet{bovy2013}, but restrict our analysis to the local neighborhood,
where the only variation is in the vertical direction. So, our MAPs
are distinguished solely by their different velocity dispersions,
$\sigma$.

We begin in \S 2 with some preliminaries and two one-dimensional
models for a local patch of the Galaxy.  The more realistic of these
contains a stellar disc, a dark halo, and a set isothermal tracer
subpopulations, which are perturbed by a breathing mode.  In \S 3, we
analyze mock catalogs generated from this model and infer the
parameters of the underlying potential and vertical force.  This
analysis allows us to quantify the systematic errors that arise when
the tracers are not in equilibrium.  In Section \S 4, we argue that
breathing mode perturbations may lead to variations of the inferred
$K_z$ with $\sigma$, which may be detectable with data from the Gaia
mission \citep{perryman2001, lindegren2008}.

\section{Perturbations in a Localized Patch of the Galactic Disc}

In this section, we describe one-dimensional, plane-symmetric models
for a local patch of the Galaxy.  We assume that vertical motions
decouple from motions in the disc plane and that gradients in the
plane of all physical quantities can be ignored. For a discussion of how variations of the potential and DF can affect the Oort problem, see Garbari et al 2011.  The models assume a
collisionless tracer population that responds to the gravitational
potential.

\newcommand\rhodm{\rho_{\rm DM}}

\subsection{Mathematical Preliminaries}

The tracer population is described by a DF $f\left (z,\,v,\,t\right )$
that satisfies the one-dimensional CBE
\begin{equation}
\frac{\partial f}{\partial t} + v\,\frac{\partial f}{\partial z} -
\frac{\partial \psi}{\partial z}\frac{\partial f}{\partial v} = 0
\end{equation}
where $\psi$ is determined by the dominant local constituents through
the Poisson equation
\begin{equation}
\frac{\partial^2 \psi}{\partial z^2} = 4\pi G\,\rho~.
\end{equation}
For a plane-symmetric equilibrium system, all quantities are time-independent and
symmetric under $z\to -z$.  In addition, $f$ is a function solely of
the vertical energy $E= v^2/2 + \psi(z)$.  Tracers follow closed
orbits in the $\left (z,\,v\right )$-plane with period $T(E)$ and
angular velocity $\omega(E) = 2\pi/T(E)$.  We can then replace $z$ and
$v$ by the canonical coordinates $E$ and $\theta$ where $d\theta =
\omega(E) dt$.  In general, it is simpler to introduce and analyze
perturbations in terms of these coordinates \citep{mathur1990,
  weinberg1991}.

For a system close to equilibrium, we can write $f = f_{0}\left
  (E\right ) + f_{1}\left (t,E,\,\theta\right )$.  Likewise, the
gravitational potential is perturbed to $\psi(z,\,t) = \psi_0(z) +
\psi_1\left (z,\,t\right )$.  In terms of $E$ and $\theta$, the CBE
becomes
\begin{equation}
\label{eq:perturbedCBE}
\frac{\partial f_{1}}{\partial t} + \omega(E)\frac{\partial
  f_{1}}{\partial \theta} - \omega(E) 
\frac{\partial \psi_1}{\partial \theta} 
\frac{df_{0}}{dE}
= 0~.
\end{equation}
The functions $f_1$ and $\psi_1$ can be written as Fourier series in
$\theta$ \citep{mathur1990, weinberg1991}:
\begin{equation}\label{eq:DFtransform}
\widetilde{f}\left (E,\,\theta\right )
= \sum_m \widetilde{f}_m\left (E\right )
e^{im\theta}
\end{equation}
and
\begin{equation}
\widetilde{\psi}_1\left (z\left (E,\,\theta\right )\right ) = \sum_m
\widetilde{\psi}_m\left (E\right )e^{im\theta}~.
\end{equation}
Doing so leads to a simple physical interpretation for the perturbed
system.  For example, the $m=1$ terms correspond to a local bending of
the disc and oscillations in $\langle z\rangle$ and $\langle
v\rangle$.  Likewise, the $m=2$ terms correspond to localized
compression and expansion and oscillations in $\langle z^2\rangle$,
$\langle v^2\rangle$, and $\langle zv\rangle$.  The latter are the 
breathing modes considered in this paper.

\citet{mathur1990}, \citet{weinberg1991}, \citet{widrow2015}
considered self-gravitating systems in which the density that appears
on the right-hand side of the Poisson equation was given by the
integral of the distribution function over velocities. They found that
the system could support true linear modes as well as Landau-damped
perturbations.  In the next subsection, we consider the mathematically
simpler problem of a system of massless tracers responding to an
external, time-dependent perturbation.  In \S 2.3, we allow for a
system of stars that both respond to and generate the time-dependent
potential.

\subsection{Two-Component Model}

For our first example we consider a two-component model where
spatially homogeneous matter (here the dark matter) generates the
potential and a single tracer responds to it. This model is simple
enough that it can be analyzed analytically and many of the lessons
learned carry over to the more complex model in the next subsection.

The  matter distribution is assumed to depend on time leading to a
potential
\begin{equation}\label{eq:potpert}
\psi\left (z,\,t\right ) = 2\pi G \rhodm \left (1 + \Delta(t)\right
  )z^2
\end{equation}
where $\rhodm$ is a constant.  The $z^2$ dependence is fairly robust
since it arises as the leading term in the Taylor expansion of a
general axisymmetric potential $\psi(R,z)$ under the assumptions that
$\psi$ and its first derivatives are continuous and that variations in
$R$ are small compared to those in $z$.  As we will see, the
$z^2$-dependence induces a breathing mode perturbation in the stellar
DF.

For illustrative purposes, we assume $\Delta(t) = \lambda \cos{\Omega
  t}$.  In the unperturbed case ($\lambda = 0$) all particles have the
same period $T = 2\pi/\omega = \pi^{1/2}/\left (G\rhodm\right
)^{1/2}$.  (For reference, the total density in the solar
neighbourhood is $\sim 0.1\,M_\odot\,{\rm pc}^{-3}$, which implies a
vertical oscillation period for stars near the midplane of $\sim 85\,{\rm
  Myr}$.)  The transformation between $\left (z,\,v\right )$ and
$\left (E,\,\theta\right )$ is then given by
\begin{equation}
z = \left (\frac{2E}{\omega^2}\right )^{1/2}\cos{\theta}~~~~~~~~
v = -\left (2E\right )^{1/2}\sin{\theta}
\end{equation}
and we can write the potential perturbation in Eq.\,\ref{eq:potpert} as $\psi_1 =
2\pi G\lambda\rhodm z^2\cos{\Omega t} = \lambda
E\cos^2{\theta}\cos{\Omega t}$.  Thus
\begin{equation}
\frac{\partial\psi_1}{\partial\theta} = \frac{i\lambda E}{2}
\left (e^{i\left (2\theta + \Omega t\right )} + 
e^{i\left (2\theta - \Omega t\right )}\right )~,
\end{equation}
which suggests the ansatz
\begin{equation}
f_1\left (E,\,\theta,\,t\right ) ~=~ 
f_+e^{i\left (2\theta + \Omega t\right )} + 
f_-e^{i\left (2\theta - \Omega t\right )}~,
\end{equation}
where it is understood that we take the real part in these expressions.
From Eq.\,\ref{eq:perturbedCBE} we find
\begin{equation}
f_1\left (E,\,\theta,\,t\right ) ~=~ \frac{\lambda\omega}{2}
\frac{df_0}{d\ln{E}} 
\left (
\frac{e^{i(2\theta + \Omega t)}}{2\omega + \Omega}+
\frac{e^{i(2\theta - \Omega t)}}{2\omega - \Omega}\right )
\end{equation}
Finally, after some algebra, we have
\begin{equation}\label{eq:DFpert}
f_1\left (E,\,\theta,\,t\right ) ~=~ 
\epsilon\left (t\right )
\frac{df_0}{d\ln{E}} 
\cos{\left (2\theta - \gamma\left (t\right )\right )}
\end{equation}
where $\alpha\equiv \Omega/2\omega$,
$\gamma = {\rm arctan}\left (\alpha\tan{\Omega t}\right )$, and
\begin{equation}
\epsilon(t) = \frac{\lambda}{2} \frac{\left (\cos^2{\Omega t} +
  \alpha^2\sin^2{\Omega t}\right )^{1/2}} {1 - \alpha^2}~.
\end{equation}

Consider a sample of $N$ tracer stars with measured phase space
coordinates $\{z_i,\,v_i\}$.  For definiteness, we assume that the
equilibrium tracer population is isothermal with DF
\begin{equation}
f_0(E) = \frac{\omega}{2\pi\sigma^2} e^{-E/\sigma^2}~.
\end{equation}
A hypothetical observer who models these stars as an equilibrium
distribution with $f_{\rm model} = f_0$ will calculate the
log-likelihood function to be

\begin{equation}\label{eq:lnlike}
\begin{split}
\ln{\cal L} & = \sum_i \ln f_0(z_i,\,v_i)\\ & = N\ln\left
(\omega/2\pi\sigma^2\right ) -\frac{1}{2\sigma^2}\sum_i \left (
\omega^2 z_i^2 + v_i^2 \right )
\end{split}
\end{equation}

\noindent The observer therefore calculates the best-fit values of
$\sigma^2$ and $\omega$ by maximizing the likelihood. Carrying out the
derivatives with respect to $\sigma^2$ and $\omega$, setting both
equal to zero, and solving the two coupled equations leads to the
estimators for the velocity dispersion
\begin{equation}
\hat\sigma^2= \langle v^2\rangle \equiv \frac{1}{N}\sum_i v_i^2
\end{equation}
and the frequency
\begin{equation}
\hat\omega^2= \frac{\hat\sigma^2}{\langle z^2\rangle}.
\end{equation}
Therefore, the estimator for the density would be 
\begin{equation}
\widehat\rhodm = \frac{\langle v^2\rangle}{4\pi G\langle
  z^2\rangle}~.
\end{equation}
Eqs. 17, 18, and 19 are of course incorrect since the data is not
described by the model. The true relationships between the model
parameters and ensemble averages are
\begin{equation}
  \langle v^2\rangle = \sigma^2\left (1 - \epsilon \cos{\gamma}\right )
\end{equation}
 and
\begin{equation}
  \langle z^2\rangle = \frac{\sigma^2}{\omega^2}\left (1 + \epsilon
  \cos{\gamma}\right )~.
\end{equation}
Meanwhile, $\langle zv\rangle =
\epsilon\sigma^2\omega^{-1}\sin{\gamma}$.

The inferred value of the dark matter density will
differ from the true one by a factor 
\begin{equation}
  \frac{\Delta\rhodm}{\rhodm} ~=~ - 2\epsilon\cos{\gamma} + O\left
  (\epsilon^2\right )
\end{equation}
As expected, the error is of order the amplitude of the
perturbation. Note that this is actually an under-estimate for how
poorly the dark matter density can be recovered. In more realistic
models, the dark matter is one of several components that contributes
to the potential and the inference about dark matter density is even
less secure.

\subsection{Three-Component Model}

We now introduce a more realistic model that will serve as a testing
bed for the analyses in subsequent sections. Here, the components are:
\begin{itemize}
\item {\bf Dark Matter:} This maintains a fixed profile contributing a
  factor proportional to $z^2$ in the potential.
\item {\bf Stellar Disc:} The unperturbed density is taken to be
\begin{equation}
  \rho_b(z) = \frac{h^2\Sigma_b}{2\left (z^2 + h^2\right )^{3/2}}
\end{equation}
where $\Sigma_b$ is the surface density. This component also
contributes to the potential and is perturbed when the potential is
perturbed.
\item {\bf Tracers:} This component comprises a series of isothermal
  stellar subpopulations distinguished by their velocity dispersion
  $\sigma$.  They participate in the perturbation but do not
  contribute to the potential.
\end{itemize}
The total equilibrium
gravitational potential is therefore
\begin{equation}\label{eq:kgpotential}
\psi(z) = 2\pi G\Sigma_b\left (\left (z^2 + h^2\right )^{1/2} - h\right )
+ 2\pi G\rhodm z^2.
\end{equation}

\begin{figure}\label{fig:phspaceDF}
\includegraphics[width=\columnwidth]{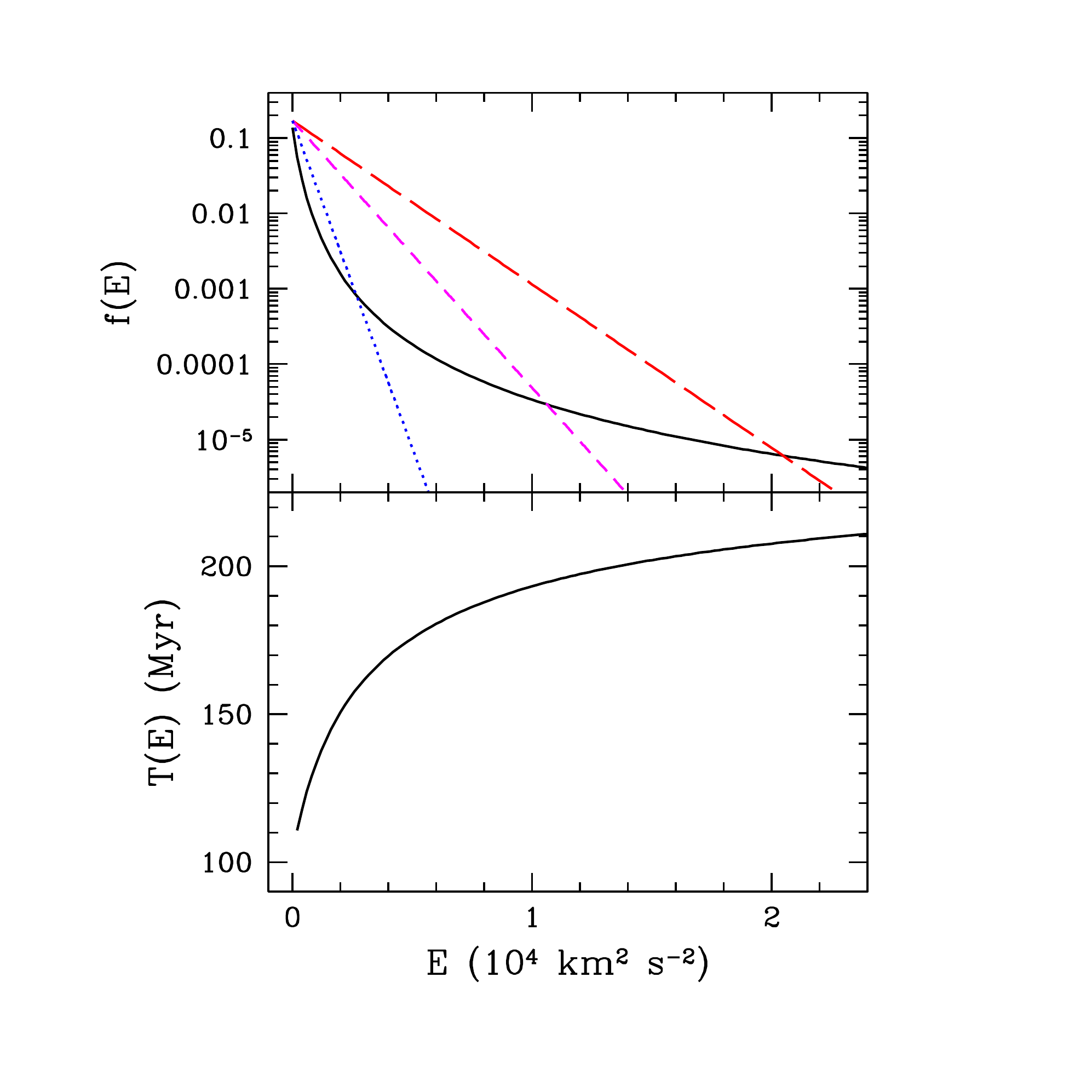}
\caption{Phase space DFs and vertical oscillation period $T(E)$ for
  the equilibrium models described in Section 2.3.  The upper panel
  shows the DFs for the disc stars (solid black), and tracer
  populations with $\sigma = 20\, {\rm km\,s}^{-1}$ (dotted blue),
  $35\,{\rm km\,s}^{-1}$ (dashed magenta) and $50\,{\rm km\,s}^{-1}$
  (long-dashed red).  The normalization of the tracer DFs is
  arbitrary and for comparison purposes, we've set it to match that of
  the disc stars at $E=0$.  The lower panel shows the vertical
  oscillation period as a function of energy.}
\end{figure}

\begin{figure}\label{fig:rhovdisp}
\includegraphics[width=\columnwidth]{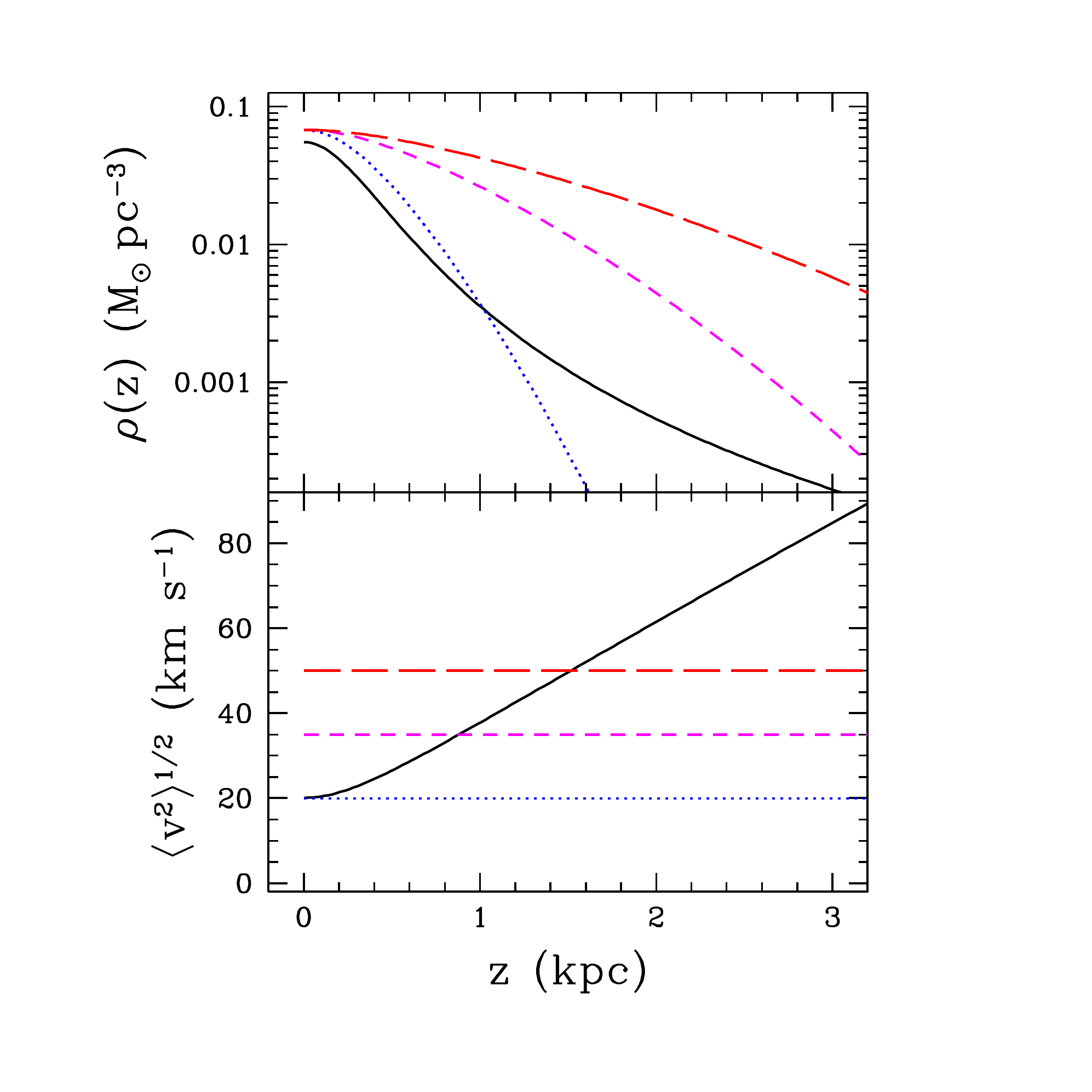}
\caption{Density and vertical velocity dispersion as a function of $z$
  for the equilibrium model described in Section 2.3.  The upper panel
  shows the vertical density profile for the stars and the tracer
  subpopulations with $\sigma = 20,\,35,\,50\,{\rm km\,s}^{-1}$.  The
  lower panel shows the velocity dispersion.  Line types and colours
  are the same as in Figure 1.}
\end{figure}

This form for the potential was introduced by \citet{kuijken1989a,
  kuijken1989b, kuijken1989c} in their series of papers on the Oort
problem.  We choose parameters to roughly match the stellar density in
the solar neighborhood.  In particular, we set $\rhodm = 0.0114\,{\rm 
  M}_\odot\,{\rm pc}^{-3}$, $\Sigma_b = 48.4\,M_\odot\,\rm{pc}^{-2}$ and $h
= 0.435\,{\rm kpc}$, which yields a vertical density profile in good
agreement with the vertical density profile from \citet{juric2008}. 
Note that ratio of the contribution to the vertical force from the
dark matter to that of the baryons from the stellar disc is $K_{DM}(z)/K_b(z)
= \left (2\rho_{DM}h/\Sigma_b\right )\left (1 + z^2/h^2\right
)^{1/2}$, which implies that at $z = 1.1\,{\rm kpc}$, the dark matter
accounts for roughly $36\,\%$ of the total vertical force.  Thus, a
$10\,\%$ systematic error in $K_{1.1}$ would imply a $25\,\%$ error in
the inferred dark matter density.

In Appendix A, we derive an analytic expression for the distribution
function of isothermal tracers embedded in this zero order potential.
Figure 1 shows the equilibrium DFs for the stellar component and for
three of the tracer subpopulations.  Note that while the tracer DFs
decrease exponentially with $E$, the DF for the stellar disc decreases
as a power-law with $E$ as $E\to \infty$, a result of the power-law
decrease in the density profile at large $z$ (See Appendix A).  Also
shown is the vertical oscillation period, which increases from $\sim
100\,{\rm Myr}$ near to the midplane, to $200\,{\rm Myr}$ at $|z|\simeq
2\,{\rm kpc}$.  Figure 2 shows the vertical density and velocity
dispersion profiles for the three stellar systems.  In principle, the
disc stars could be represented as a superposition of isothermal
populations.

We then assume that the DFs of the tracers and the stellar disc are
perturbed as in Eq.~\ref{eq:DFpert}, with $\epsilon = 0.2$ and $\gamma
= \pi/2$.  With these choices $\langle z^2\rangle$ and $\langle
v^2\rangle$ are initially equal to their equilibrium values while
$\langle zv\rangle \simeq 2-5\,{\rm km\,s}^{-1}$.  The latter is
consistent with measurements of the bulk vertical velocities in the
solar neighbourhood \citep{widrow2012, williams2013, carlin2013}.
This perturbation then feeds back into the potential of the stellar
disc The system is evolved using an N-body code in which the disc
stars and tracers are modeled as plane symmetric sheets
(one-dimensional ``particles'') that interact via gravity (see, for
example, \citet{weinberg1991}).  Gravity in a plane-symmetric system
is particularly simple since the force on a given particle at position
$z'$ is proportional to the difference between the number of particles
with $z>z'$ and the number with $z<z'$.  Thus, forces at each timestep
can be obtained by sorting the particles in $z$.
 
The stellar disc is modeled with $4\times 10^5$ particles. This number of
    particles is more than adequate for modeling one spatial and two
    phase space dimensions as test simulations with fewer disc
    particles confirm. For the
tracers, we note that the stellar surface density at the position of
the Sun is $\sim 50\,\rm{M_\odot \,pc^{-2}}$, which implies that a
local patch of the disc $1\,{\rm kpc}$ across will contain some
$10^7-10^8$ stars.  The Gaia mission \citep{perryman2001,
  lindegren2008} aims to provide kinematic data for a large fraction
of these stars.  One might then imagine dividing these stars into
$O(100)$ subpopulations that are defined by chemical abundances as in
\citet{bovy2013}.  Each of these subpopulations can then be used as an
independent, isothermal tracer of the gravitational potential.  With
these numbers in mind, we model each of the tracer subpopulations with
$10^5$ particles.

In Figure 3 we show the evolution of $\langle z^2\rangle$, $\langle
v^2\rangle$, and $\langle zv\rangle$ in the presence of the
perturbation.  The general features of the oscillations are easy to
understand.  First, the vertical oscillations for the coldest tracers
($\sigma = 20\,{\rm km\,s}^{-1}$) have a period $\sim 70\,{\rm Myr}$.
As expected for an $m=2$ breathing mode, this is half the vertical
oscillation period for a typical star in this subpopulation (see
Figure 1).  The oscillation periods for the $\sigma = 35\,{\rm
  km\,s}^{-1}$ and $50\,{\rm km\,s}^{-1}$ subpopulations are somewhat
longer, consistent with the fact that these populations are comprised
of stars with higher vertical energies and therefore longer
oscillation periods.  The oscillations damp due to phase mixing and
the damping is strongest for the coldest population where the
dynamical time is shortest.  We conclude that {\it the amplitude and
  phase of vertical oscillations in different subpopulations need not
  be the same.}

\begin{figure}\label{fig:timeevol}
\includegraphics[width=\columnwidth]{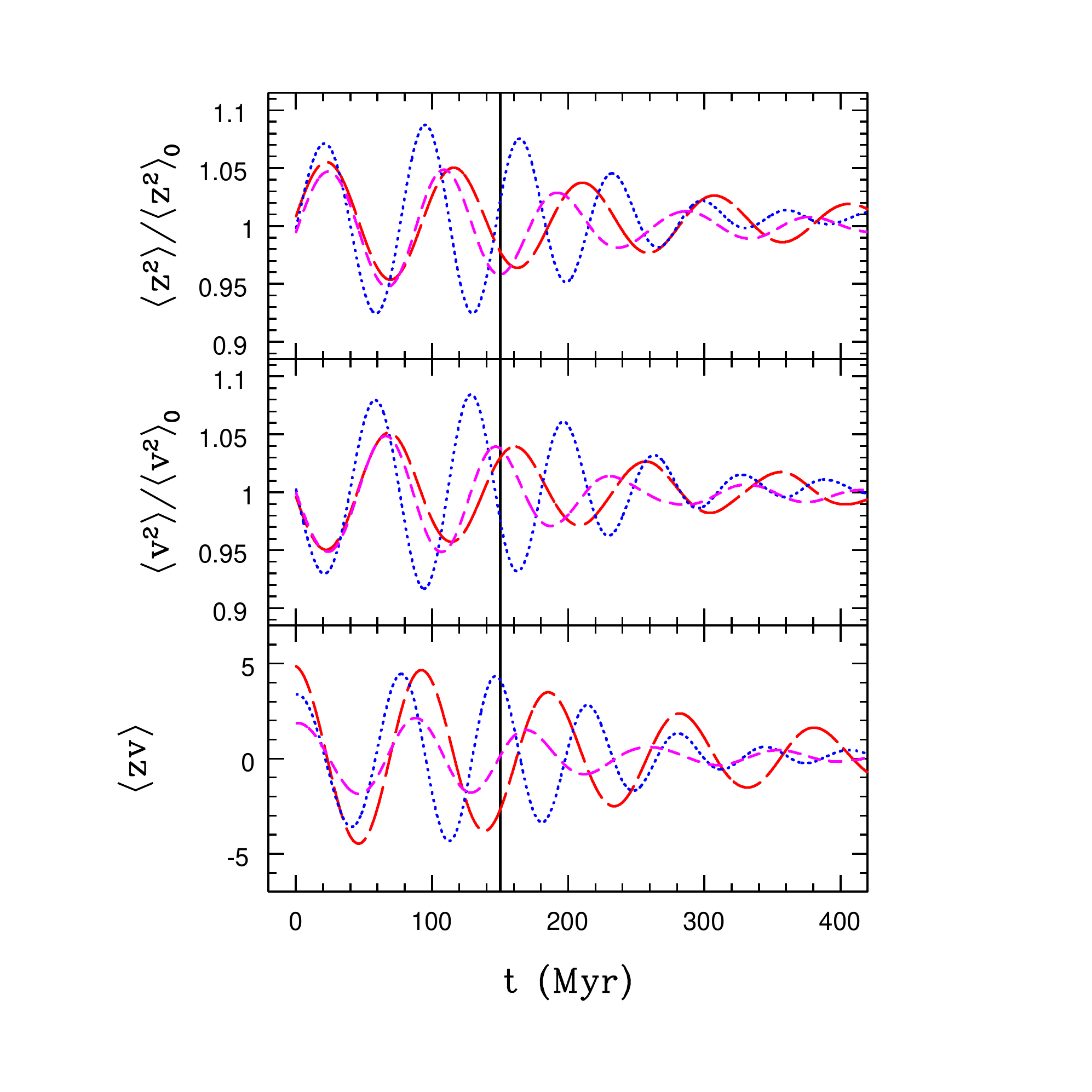}
\caption{Time evolution of the variance in $z$ and $v$ as well as $\langle
  zv\rangle$ for three subpopulations.  The top panel shows the
  variance in $z$ normalized to the equilibrium value for $\sigma =
  20,\,35,\,50\,{\rm km\,s}^{-1}$.  Line types and colours are the same
  as in Figure 1.  The middle panel shows the same for $v$.  Bottom
  panel shows the time evolution of $\langle zv\rangle$ in units of
  ${\rm km\,s}^{-1}\,{\rm kpc}$.  The vertical line here and in Figure
  4 indicates the epoch at which we generate the mock catalogs that
  are analyzed in \S 3.}
\end{figure}

In Figure 4 we show the time evolution of the surface density within
$1.1\,{\rm kpc}$ as well as an estimator for the vertical force
$\langle v^2\rangle /\sqrt\langle z^2\rangle$.  We see that the
amplitude of the oscillations in the former are an order of magnitude
smaller than those of the latter.  Thus errors in estimates of the
local vertical force that arise when the tracers are out of
equilibrium are likely to be far more significant than oscillations in
the force itself.

\begin{figure}\label{fig:timeevol2}
\includegraphics[width=\columnwidth]{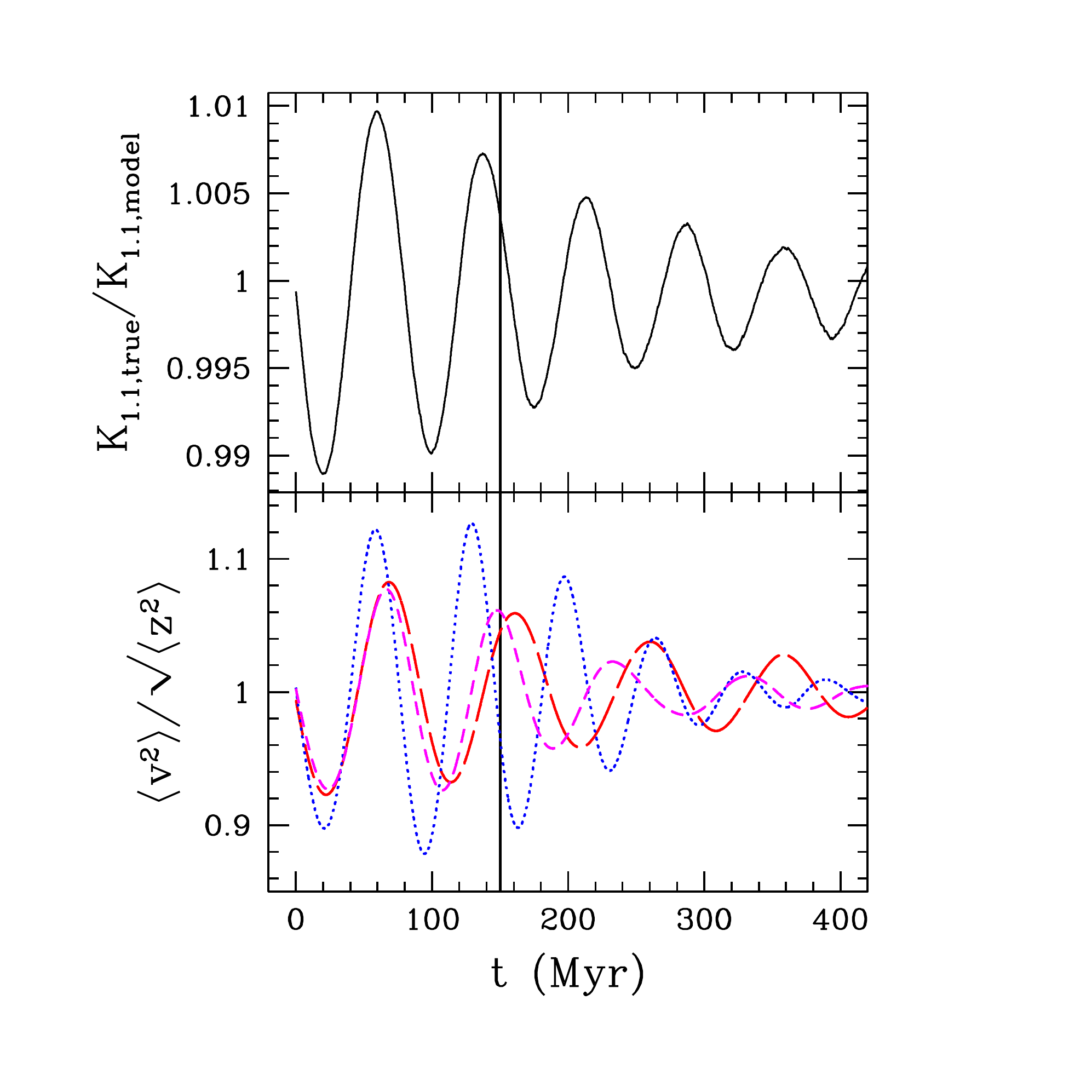}
\caption{In the presence of a perturbation, the time evolution of the
  surface density (top) and an estimator for it, $\langle
  v^2\rangle/\sqrt{\langle z^2\rangle}$.  The latter is an estimator
  for the vertical force.  Both quantities have been normalized to
  their equilibrium values. Line types and colours are the same as in
  Figure 3.}
\end{figure}

\section{parameter biases induced by non-equilibrium effects}

In this section, we treat kinematic snapshots of the simulation
described in \S 2.3 as mock data that can be analysed to infer the 
gravitational potential and force.  The results are then compared with
the equilibrium potential and force and the true (that is, perturbed)
potential and force.

For a sample of $N$ tracers from a particular subpopulation, the assumed
likelihood function is 
\begin{equation}\label{eq:likelihood}
\mathcal{L}\left (\rhodm,h,\Sigma_b,\sigma\right ) = \prod_{i=1}^N 
\left[{\cal N}(\rhodm,h,\Sigma_b,\sigma)\, 
e^{-E_i/\sigma^2}\right]
\end{equation}
where $E_i = v_i^2/2 + \psi(z_i)$ is the energy of the $i^{\rm{th}}$ star
from the sample,$\psi$ is given by Eq.\,\ref{eq:kgpotential},
and
\begin{equation}
{\cal N} = \left (\int dz\,dv \exp{\left (-E/\sigma^2\right )} \right
)^{-1}
\end{equation}
is a normalization constant.  We assume uniform priors in the model
parameters $\rhodm , h,\,\Sigma_b$ and $\sigma$ and sample the
posterior probability distribution function (PDF) using EMCEE
\citet{foreman2013}, which implements the ensemble sampler of
\citet{goodman2010}.

Here we investigate the extent to which the equilibrium assumption
will {\it bias} the parameters of the potentials, that is, the amount by
which the parameters are mis-estimated when the true distribution has a
non-equilibrium signature as described in \S 2.

Figure 5 shows the model PDFs that are inferred from two mock data
sets for the $\sigma = 35\,{\rm km\,s}^{-1}$ tracer subpopulation.
The first data set is drawn from an equilibrium distribution while the
second is drawn from the $150\,{\rm Myr}$ snapshot of the simulation.
As expected, when the data is drawn from an equilibrium distribution,
that is, when the model correctly describes the data, the analysis
recovers the model parameters to within the calculated uncertainties.
We note that there is a strong negative correlation between $\Sigma_b$
and $\rhodm$, which indicates a degeneracy in the disc and halo
contributions to the potential.  That is, the data are most sensitive
to the total force and this can be kept close to fixed by increasing
the dark matter density while decreasing $\Sigma_b$. The
    degeneracy between dark and visible matter was noted
    previously by Bahcall (1984a), Kuijken \& Gilmore (1991), and
    Garbari et al. (2012) (see also Read 2014) and can be partially
    broken by extending the sample of stars with larger values of $z$.  There is also a
strong positive correlation between $\Sigma_b$ and $h$, which we might
have anticipated by considering the leading term in the Taylor
expansion of the disc contribution to the potential: $2\pi G
\Sigma_bz^2/2h$.

The striking result in Figure 5 is that inferred value of the local
dark matter density differs by a factor of two from the true value in
the presence of this (quite realistic) breathing mode.  Not
surprisingly, when viewed in the $\Sigma_b-\rhodm$ and $\Sigma_b-h$
planes, these departures tend to lie along the degeneracies mentioned
above.  Thus, we expect that the inferrences in the vertical force or,
alternatively, the total surface density will be more robust.  This
point was discussed in \citet{kuijken1991} and Figure 6 shows that it
is indeed the case.  In particular, when the $35\,{\rm km\,s}^{-1}$
tracers are perturbed, $K_{1.1}$ is over-estimated by only about
$10\%$.  With a sample size of $10^5$ stars and ``perfect'' data (we
have made no attempt to model observational uncertainties) this
systematic error still represents a $5$-sigma departure from the true
value.

\begin{figure}\label{fig:sig35corner}
\includegraphics[width=\columnwidth]{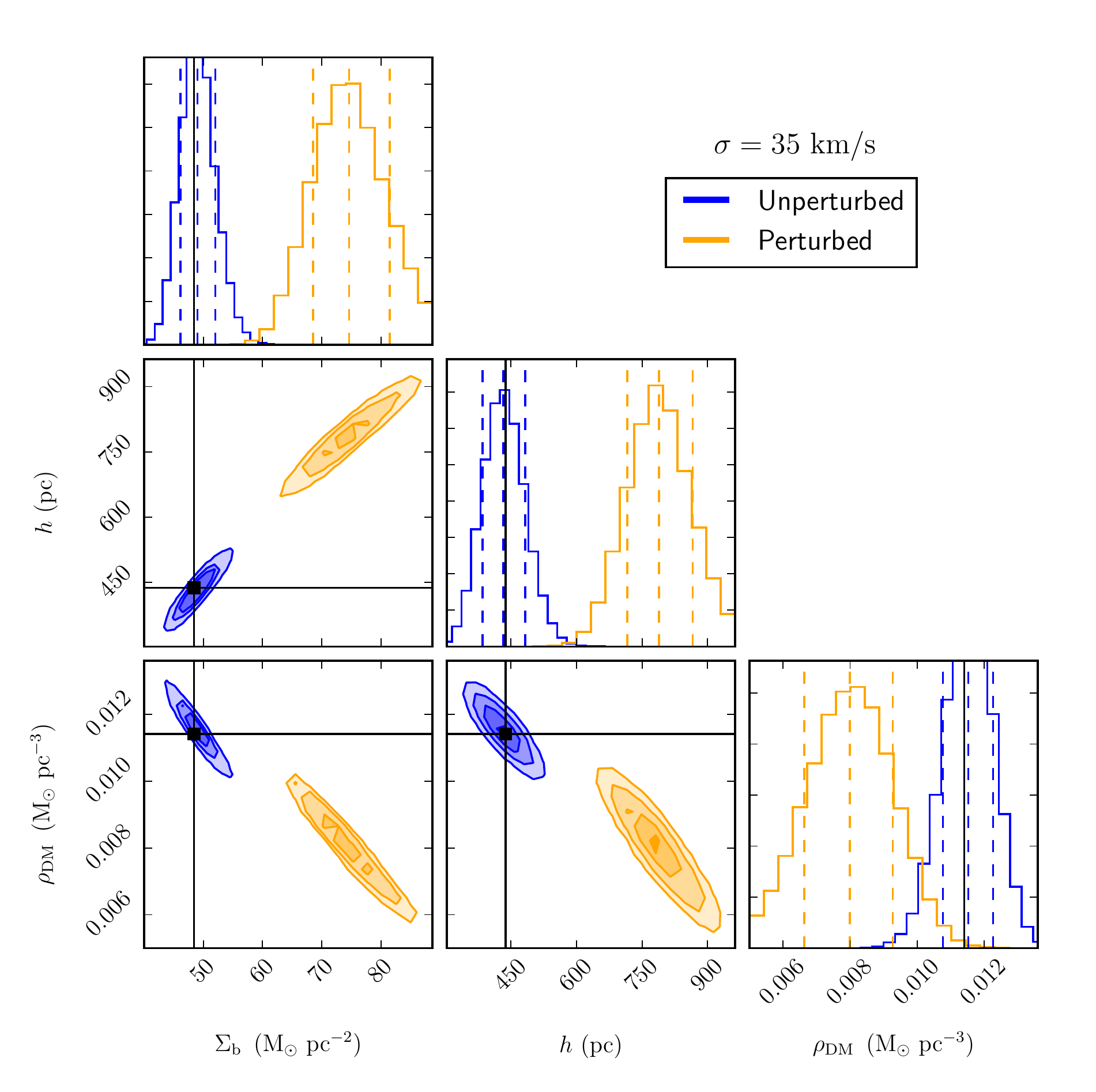}
\caption{PDF in the model parameter space given two mock data sets
  that sample the $\sigma=35\,{\rm km\,s}^{-1}$ subpopulation.  Each
  panel shows a different two-dimensional projection of the PDF in the
  parameter space defined by $\left (\Sigma_b,\, h,\,
  \rhodm,\,\sigma\right )$.  Orange contours are for a sample drawn
  from the equilibrium distribution; blue contours are for a sample
  drawn from a distribution that has been perturbed by a breathing
  mode, namely the $150\,{\rm Myr}$ snapshot of the N-body simulation
  described in \S 2.3.}
\end{figure}

\begin{figure}\label{fig:sig35force}
\includegraphics[width=\columnwidth]{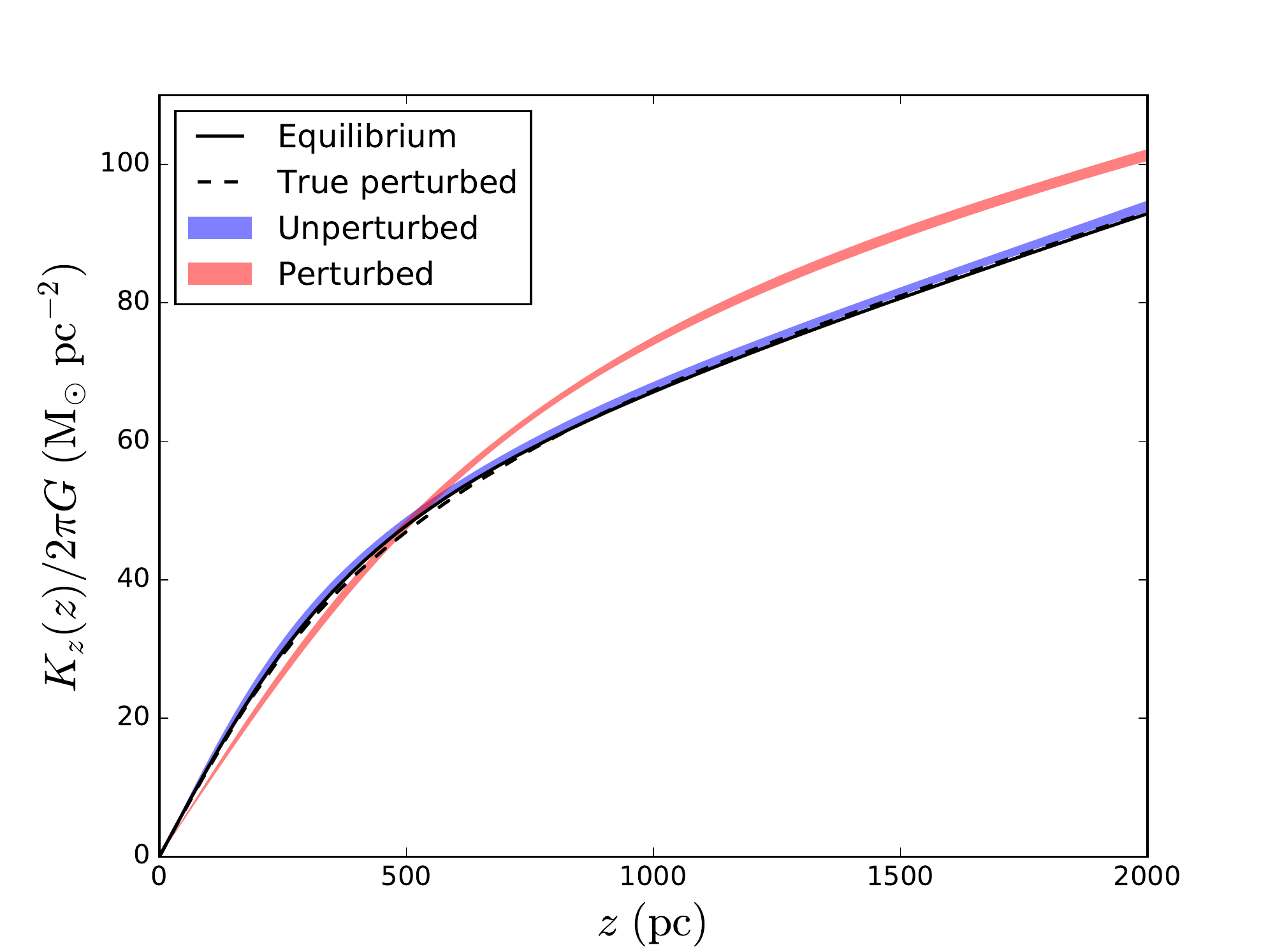}
\caption{Magnitude of the vertical force $K_z(z)$ as inferred from the
  mock data sets described in \S 3.2 and used in Figure 5.  Bands show
  the 68\% confidence intervals for $K_z$ as a function of $z$.  The
  blue band is for the equilibrium sample while the orange band is for
  the sample perturbed by a breathing mode.  Also shown is the
  equilibrium vertical force (solid line) and true perturbed vertical
  force (dashed line).}
\end{figure}

\section{From the Vertical Force to Disk Perturbations}

It is an implicit assumption in the Oort problem that different tracer
subpopulations will infer the same vertical force $K_z(z)$ to within
the calculated uncertainties.  This assumption is greatly exploited in the
analysis of \citet{bovy2013} where dozens of MAPs are used as
independent tracers of the gravitational potential.  In this section
we argue the converse: differences in the force inferred from
different tracer subpopulations may provide evidence that the disc
is in a perturbed state.

We begin by re-examining the results from \citet{bovy2013}.  Their
analysis was based on a sample of 16K G dwarfs from SEGUE
\citep{yanny2009} separated into 43 MAPs.  For our purposes, each MAP
can be distinguished by its vertical velocity dispersion $\sigma$ and
a characteristic Galactocentric radius $R$.  Our contention is that
breathing mode perturbations may induce a dependence of $K_z$ on
$\sigma$ for subpopulations at the same $R$ though as we'll see, the
precise nature of this dependence cannot be known {\it a priori}.

\citet{bovy2013} find that the MAPs with higher $\sigma$ tend to
be closer to the Galactic centre.  Furthermore, the vertical force at
fixed $|z|$ decreases with increasing $R$.  In particular
\citet{bovy2013} find that vertical force at $|z|=1.1\,{\rm kpc}$ is
well-fit by the exponential
\begin{equation}\label{eq:expdisc}
\frac{K_{1.1}(R)}{2\pi G} = 
67\,M_\odot {\rm pc}^{-2} 
\exp{\left (-\left (R-R_0\right )/2.7\,{\rm kpc}\right )}
\end{equation}
where $R_0 = 8\,{\rm kpc}$ is the distance of the Sun from the
Galactic centre.  Together, these results imply that there is an
``accidental'' correlation between $\sigma$ and $K_{1.1}$.  To remove
this correlation we correct $K_{1.1}$ using Eq.\,\ref{eq:expdisc} so
that each MAP provides an estimate of the vertical force at $R=R_0$.
In addition, we separately consider estimates for $K_{1.1}$ from
subpopulations that probe the potential within $1$ kpc bands in $R$.
The results for $6\,{\rm kpc}<R<7\,{\rm kpc}$ and $7\,{\rm
  kpc}<R<8\,{\rm kpc}$ are shown in Figure 7.  These results are
consistent with the null hypothesis that $K_{1.1}$ is independent of
$\sigma$ though there are hints of a trend toward systematically
higher values of $K_{1.1}$ among the low-$\sigma$ subpopulations with
$R$ between $6$ and $7$ kpc.

\begin{figure}\label{fig:bovyrix}
\includegraphics[width=\columnwidth]{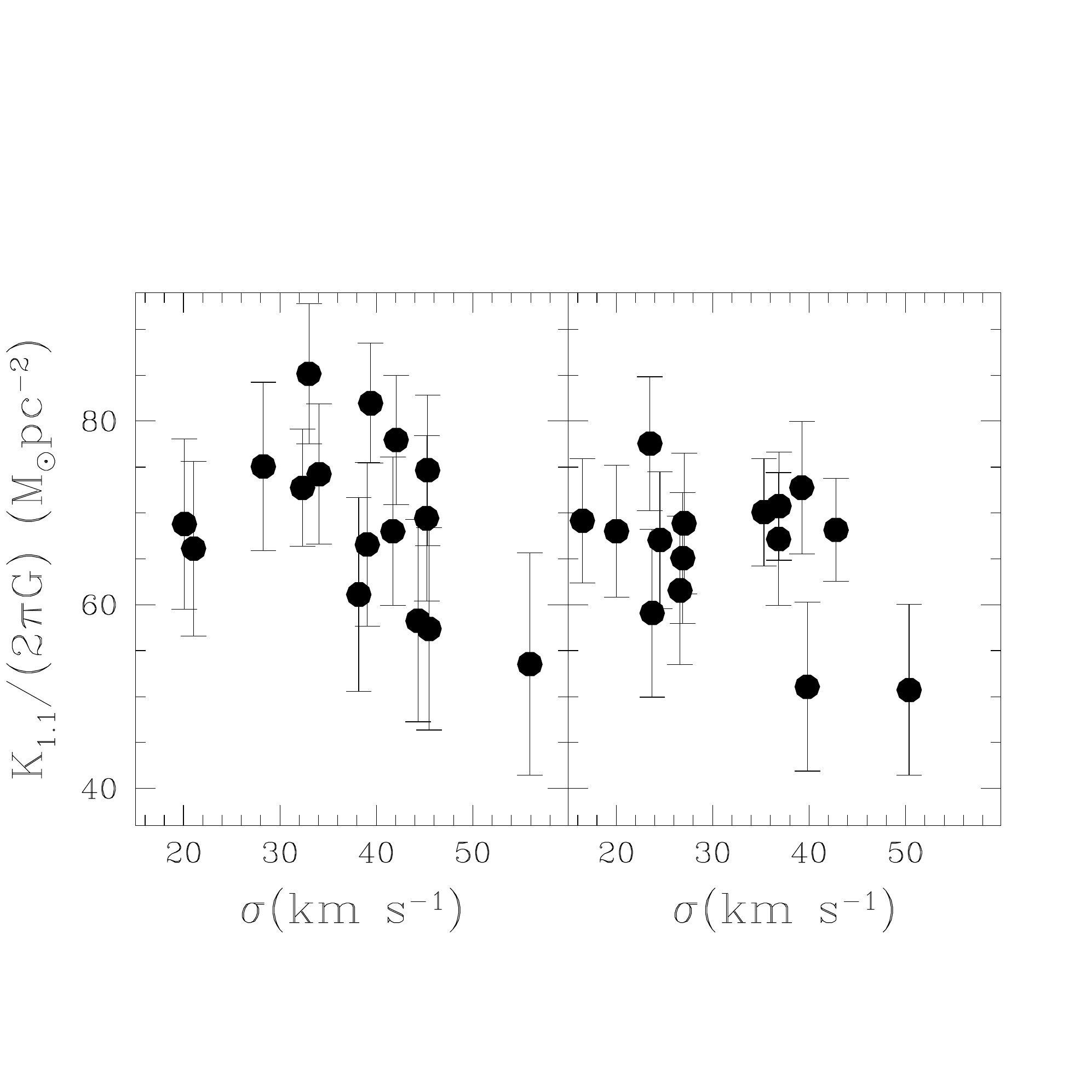}
\caption{$K_{1.1}$ vs. $\sigma$ for different MAPs from the analysis
  of \citet{bovy2013}.  Left panel shows MAPs with characteristic
  radius in the range $6 {\rm kpc} < R < 7\,{\rm kpc}$ while the right
  panel is for the range $7\, {\rm kpc} < R < 8\,{\rm kpc}$.  Values
  of $K_{1.1}$ have been corrected to the position of the Sun using
  Eq.\,\ref{eq:expdisc}.}
\end{figure}

With only 100-800 stars in each MAP, the fractional uncertainties in
$K_{1.1}$ found by \citet{bovy2013} are $10-20\%$ and therefore
comparable to the anticipated effects of a breathing mode
perturbation.  Fortunately Gaia will increase the sample size by two
or more orders of magnitude and therefore reduce the uncertainties by
a factor of 10 or greater.  With this in mind, we investigate whether
the variations in $K_z$ with $\sigma$ that are induced by a breathing
mode perturbation might be detected when the subpopulation sample size
is $10^5$.

In Figure 8 we show PDFs for the model parameters that are inferred
from the $\sigma=20\,{\rm km\,s}^{-1}$, $\sigma=35\,{\rm km\,s}^{-1}$
and $\sigma=50\,{\rm km\,s}^{-1}$ subpopulations.  The mock data
samples for these subpopulations are taken from the $150\,{\rm Myr}$
snapshot of the simulation.  We see that the parameter determinations
from the different tracers can disagree with one another at the
multiple-sigma level (e.g., the ($h,\rhodm$) plane). This disagreement
will provide a signal that the underlying model is incorrect.  As in
Figure 5, the confidence intervals from the different mock data sets
tend to line up along the correlation ridges mentioned above.
Nevertheless, there are departures off these ridges indicating that
the different data sets will lead to slightly different estimates for
$K_z$.  Figure 9 shows the force inferred from each tracer. Although
all the three inferences differ from the truth by ~20\%, they disagree
with one another at only about the 5\% level.

\begin{figure}\label{fig:3popcorner}
\includegraphics[width=\columnwidth]{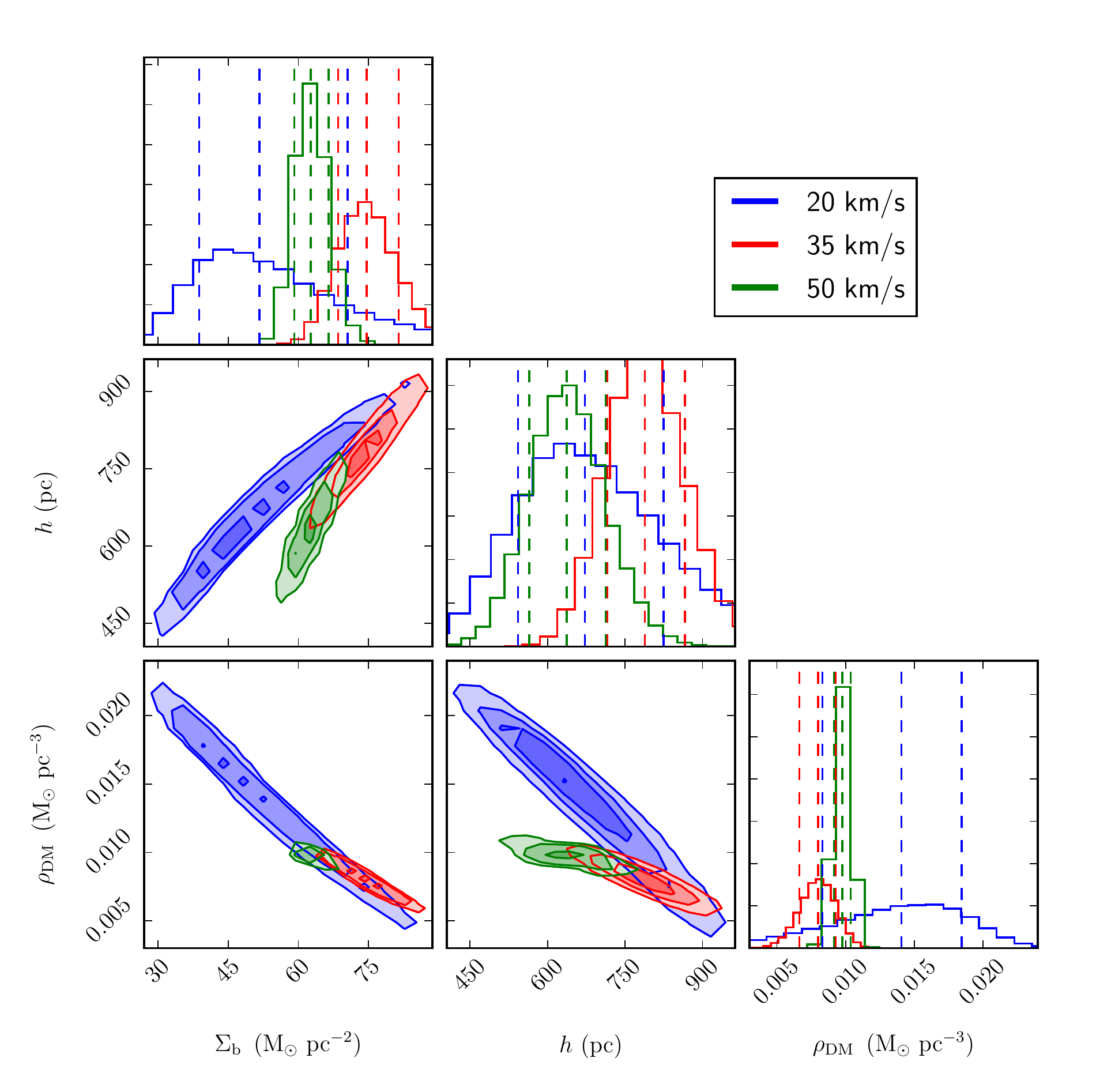}
\caption{PDF for the model parameters as inferred from three different
  perturbed subpopulations.  The mock data sets are from the
  $150\,{\rm Myr}$ snapshot of the simulation described in \S 2.3.
  Blue contours are from the $\sigma=20\,{\rm km\,s}^{-1}$ sample, red
  contours are from the $\sigma=35\,{\rm km\,s}^{-1}$ sample and
  therefore the same as the contours found in Figure 5 and green
  contours are from the $\sigma=50\,{\rm km\,s}^{-1}$ sample.}
\end{figure}

In Figure 10, we present a scatter plot of $\Sigma_b$ and $\rhodm$ as
inferred from an analysis of mock data for eight subpopulations with
$\sigma = 20, 25, 30, 35, 40, 45, 50 ~ \rm{and} ~ 55\,{\rm
  km\,s}^{-1}$ at five different snapshots of our simulation.  This
figure can be compared with the lower left panels of Figure 5 and 8.
Recall that the initial conditions were chosen so that $\langle
z^2\rangle$ and $\langle v^2\rangle$ were equal to their equilibrium
values.  Therefore, it is not surprising that with the initial
snapshot (solid black circles), the true model parameters are
recovered quite accurately.  For the later snapshots, when $\langle
z^2\rangle$ and $\langle v^2\rangle$ depart from their equilibrium
values, the different tracers can yield significantly different values
for the model parameters.  The results also vary significantly from
snapshot to snapshot, a reflection of the stochastic nature of disc
perturbations.  The models do tend to lie along a narrow ridge in
$K_{1.1}$ and this once again illustrates that it is the vertical
force or total surface density that is most robustly determined from
the stellar dynamics.

To further illustrate how $K_z$ might depend on $\sigma$ we show, in
Figure 11, $K_{1.1}$ for the subpopulations and simulation snapshots 
used in Figure 10.  Once again, for the initial conditions
the model recovers the true value of $K_{1.1}$ to within the
calculated uncertainties.  Each of the other snapshots show a
different example of what a $K_{1.1}$-$\sigma$ curve might look like.
The model might systematically overestimate $K_{1.1}$, as with the
$50$ and $150\,{\rm Myr}$ snapshots or underestimate $K_{1.1}$, as
with the $100\,{\rm Myr}$ snapshot.  Typical variations in $K_{1.1}$
across the range in $\sigma$ are $\sim 5\%$ though in one example, the
$200\,{\rm Myr}$ snapshot the variation is greater than $10\%$.

\begin{figure}\label{fig:force3pop}
\includegraphics[width=\columnwidth]{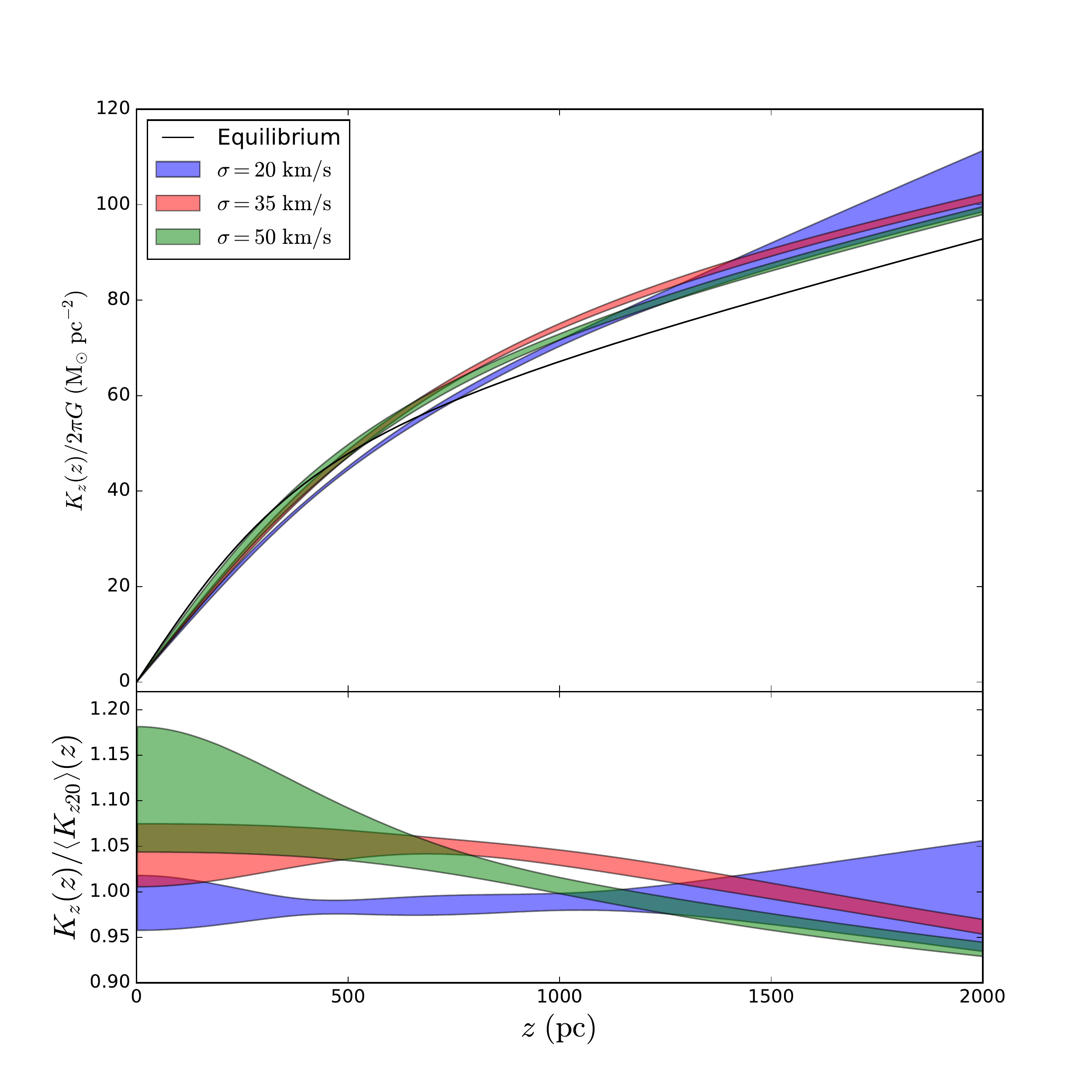}
\caption{68\% confidence intervals for $K_z$ derived from the PDFs
  shown in Figure 8.  In the lower panel, we show
  the confidence intervals for $K_z$ divided by the mean values for
  the $\sigma=20\,{\rm km\,s}^{-1}$ sample.  Also shown as a solid
  black line is $K_z$ for the equilibrium potential.}
\end{figure}

\begin{figure}\label{fig:rhovssigmab}
\includegraphics[width=\columnwidth]{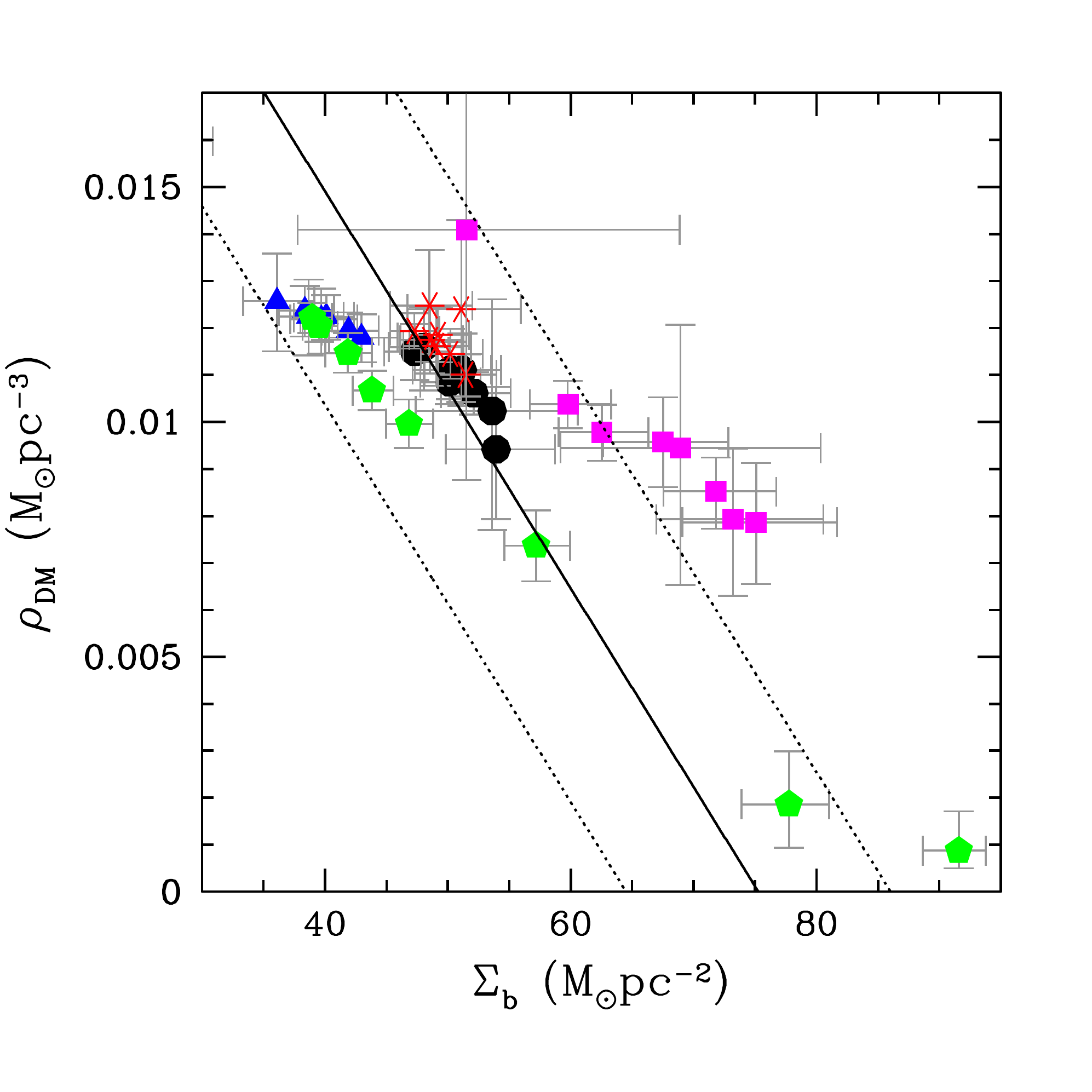}
\caption{Estimates for $\Sigma_b$ and $\rhodm$ from eight subpopulations
at five snapshots from the simulation described in \S 2.3.  Initial
conditions are shown as black circles.  Other snapshots are shown as 
red stars, blue triangles, magenta squares, green pentagons for
snapshots spaced at $50\,{\rm Myr}$ intervals.}
\end{figure}

\begin{figure}\label{fig:k11snap}
\includegraphics[width=\columnwidth]{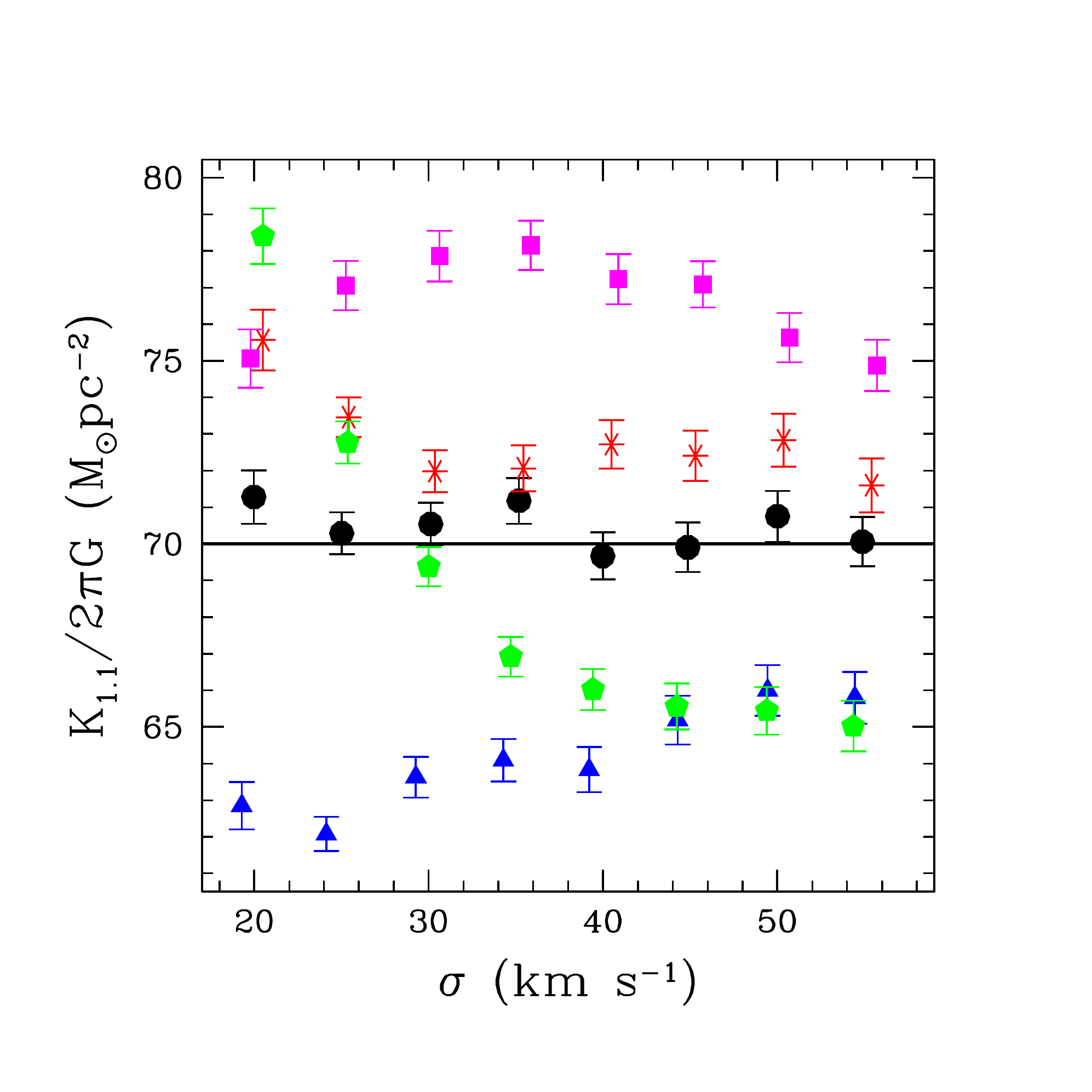}
\caption{Estimates for $K_{1.1}$ as a function of $\sigma$ for 
eight subpopulations and five snapshots.  Point types and colours are
the same as in Figure 10.}
\end{figure}

\section{Conclusions}

The traditional Oort analysis assumes an underlying equilibrium
distribution for stars in the solar neighbourhood. Perturbations in
the potential, and indeed there is evidence already for a perturbation
in the form of a breathing mode, can upset the inferences from these
analyses. In the presence of this perturbation, a mismatch between the
true and inferred values of the vertical force has important
implications, especially as we enter the Gaia era, where observations
of many different tracers, each with of order $10^5$ velocities and
positions, are feasible. Here we have run mock observations to
quantify the following three effects of a perturbed disc:
\begin{itemize}
\item The vertical force as inferred from a single tracer may differ
  from the true value by $10\%$ or greater, depending on the phase and
  amplitude of the perturbation.  This error is distinct from other
  sources of statistical and systematic uncertainties inherent in the
  Oort problem.  The corresponding error in an estimate of the local
  dark matter density would be at the $25\%$ level.
\item An analysis of multiple tracers, each with different velocity
  dispersions, will lead to inconsistent conclusions about the total
  surface density profile thereby providing evidence that the
  underlying model -- assumed to be equilibrium -- was wrong. With
  enough tracers, one could imagine discovering something about the
  cause of the perturbation, be it a passing dark matter sub-halo or
  nearby dwarf galaxy or some other transient phenomena in the disc.
\item Agreement among the conclusions from multiple tracers would
  improve the robustness of the dark matter determination: if all the
  tracers give the same answer, we can be
  confident that the underlying model and the conclusions inferred
  from it are correct. 
  
  The main limitation of our analysis is that it treats
    the Galaxy as a plane-symmetric system.  This simplification
    allowed us to focus on the effects of a perturbed disc.  In the
    full three-dimensional Galaxy, additional complications might
    arise, which could masquerade as disc perturbations.  For example,
    a tilt of the velocity ellipsoid away from the plane of the disc
    can bias estimates of the dark matter density if not properly
    modeled (see for example, Kuijken \& Gilmore (1989a,b,c), Garbari et
    al. (2012), and more recently, Silverwood et al. (2016)).  Of course, in
    a perturbed disc, the tilt might vary across the different MAPs
    and, in principle, could itself be a time-dependent feature of the
    perturbed disc.  The promise of Gaia is that we will be able to
    disentangle these different effects and ultimately place robust
    constraints on the local density of dark matter.
\end{itemize}

{\textbf{Acknowledgments} --} The authors are grateful to Brian
Yanny, Alex Drlica-Wagner, Elise Jennings, and Jo Bovy for useful
comments and discussions. Fermilab is operated by Fermi Research
Alliance, LLC, under Contract No. DE-AC02-07CH11359 with the
U.S. Department of Energy. NB was supported by the Fermilab Graduate
Student Research Program in Theoretical Physics. LMW was supported by a Discovery Grant with the Natural Sciences and Engineering Research Council of Canada.

%

\section{Appendix}

In this section, we derive the analytic DF for the baryon component introduced
in Section 2.3.  In general, the density for a plane-symmetric system is derived
from the DF by the integral $\rho(z) = \int dv f(z,\,v)$.  For an equilibrium
system

\begin{equation}
\rho\left (\psi\right ) = \frac{1}{\sqrt{2}}
\int_\psi^\infty \frac{dE f(E)}{\sqrt{E-\psi}}
\end{equation}

\noindent By an Abel transform, we have

\begin{equation}
f(E) = -\frac{1}{\sqrt{2}\pi}\int_E^\infty \frac{d\rho}{d\psi}
\frac{d\psi}{\sqrt{\psi - E}}
\end{equation}

\noindent It is convenient to write the potential and 
density in terms of $u \equiv \left (1 + z^2/h^2\right )^{1/2}$:

\begin{equation}
\psi = 2\pi G\left (\rho_{dm} h^2 u^2 + \Sigma_b h u - \Sigma_b h - \rho_{dm}
h^2\right )
\end{equation}

\noindent and

\begin{equation}
\rho_b = \frac{\Sigma_b}{2hu^3}~.
\end{equation}

We then have

\begin{align}
f(E) & = \frac{3\Sigma_b}{\sqrt{8}\pi h}
\int_{u(E)}^\infty \frac{du}{u^4\sqrt{\psi(u) - E}}\\
& = \frac{3}{4\pi}\frac{\Sigma_b}{h}
\frac{1}{\sqrt{\pi G\Sigma_b h}}
\int_{u(E)}^\infty \frac{du}{u^4\left (\alpha u^2 + u - \mathcal{E}\right )^{1/2}}
\end{align}

\noindent where $\alpha\equiv \rho_{dm}h/\Sigma_b$, 
$\mathcal{E} \equiv E/\left (2\pi G\Sigma_b h\right )
+ 1 + \alpha$ and $u(E) = \left (\left (1 + 4\alpha\mathcal{E}\right )^{1/2} - 1\right )/2\alpha$.  The integral can be expressed in terms of elementary functions
are we have

\begin{equation}
f(E) = 
\frac{3}{4}\left(
\frac{\Sigma_b}
{\pi^3 G h^3}\right )^{1/2}
\frac{1}{\mathcal{E}^{7/2}}
\mathcal{F}\left (\left (\alpha\mathcal{E}\right )^{1/2}\right )
\end{equation}

\noindent where

\begin{equation}
\mathcal{F}(x) = \frac{1}{96}\left (
\left (72x^2 + 30\right ){\rm ctn}^{-1}(2x)
+ 15\pi + 60x + 36\pi x^2 + 64x^3\right )~.
\end{equation}

\label{lastpage}

\end{document}